\documentclass[reprint,twocolumn,
amsmath,amssymb,aps,
prd
]{revtex4-2}
\usepackage{graphicx,subfigure,color}
\usepackage[caption=false]{subfig} 
\usepackage{amssymb}
\usepackage{amsmath}
\usepackage{commath}
\usepackage{graphicx,bm}
\usepackage{verbatim}
\usepackage{graphicx}
\usepackage{dcolumn}
\usepackage{bm}

\usepackage{xcolor}

\newcommand{\be}{\begin{equation}}
	\newcommand{\ee}{\end{equation}} 
\newcommand{\ba}{\begin{array}}
	\newcommand{\ea}{\end{array}}
\newcommand{\bea}{\begin{eqnarray}}
	\newcommand{\eea}{\end{eqnarray}}

\usepackage{multirow}
\begin{document} 
	\preprint{APS/123-QED}	
	\title{  Scaling dimension of Cooper pair operator from the black hole interior}
	\author{Yoon-Seok Choun$^{a,b}$}
	\email{ychoun@gmail.com}
	\affiliation{$^{a}$Department of Physics, POSTECH, Pohang, Gyeongbuk 37673, Korea \\ $^{b}$Asia Pacific Center for Theoretical Physics (APCTP), Pohang, Gyeongbuk 37673, Korea}
	\author{Sang-Jin Sin$^{c}$}%
	\email{sangjin.sin@gmail.com}
	\affiliation{$^{c}$Department of Physics, Hanyang University, Seoul 04763, South Korea
	}%

	\date{\today}
	
	\begin{abstract}
		We have shown that in 
		holographic superconductivity theory for 3+1 dimensional system,  the scaling dimension  of Cooper pair operator can be obtained as a quantized value  if  we request that the the   scalar function describing the order parameter is finite inside the black hole  as well as outside.  This should be contrasted to the usual situation where we set the mass squared of the scalar by hand.   Our method  can be applied to any order parameters. 	
	\end{abstract}
	
	\maketitle
	
	\section{Introduction}
	
 Calculating the anomalous dimension in the interacting field theory is highly non-trivial task.
Even in holographic theory\cite{Gubser:2008px,Hartnoll:2008vx}, scaling dimension has been the  input data which  was set to be an integer $\Delta=1,2$   by hand. 
Certainly this is not desirable, because, for example, the scaling dimension of the Cooper pair operator can not be an arbitrary number, and the detailed behavior of the superconductivity depends on this number very sensitively. 

In this paper, we analyze the gap equations of holographic superconductors in 3+1 dimension and  show that 
 in the presence of the horizon, the regularity of the condensating solution inside the  black hole  provides  a simple way to calculate the scaling dimension, because  the higher order singularity  requests extra regularity in  the solution, leading to the quantized value of the scaling dimension. And we require that the solution is a polynomial after factoring out the singular pieces. Then the solution automatically satisfies the horizon regularity, which is the condition usually   imposed in the literature. 

 We analyzed analytically all the allowed spectrum in the probe limit of the background gravity near the critical temperature. The lowest possible scaling dimension is $\Delta=2$ and the next one is about 3.6.etc. 
This is analogous to the energy quantization in Schroedinger equation.  	The generality of our method  comes from the ubiquitous appearance  of the Heun's equation  in the holographic setup of symmetry breaking regardless of the spin of the matter fields or dimension of the bulk spacetime\cite{Gubser:2008px,Hartnoll:2008vx,Gubser:2008wv,benini2011holographic}.  

	\vskip.5cm
	\section{Set up}
	{	We consider the action \cite{Hart2008},} 
	\begin{small}
		\be  {
			S=\int d^{d+1}x \sqrt{-|g|}\left( -\frac14 F_{\mu\nu}^{2}- |D_{\mu}\Psi |^{2}-m^{2}|\Psi|^{2} \right),}
		\ee
	\end{small}
	where $|g|=\det g_{ij}$,   $D_{\mu}\Psi=\partial_{\mu}-igA_{\mu}$ and $F=dA$, and $A=\Phi dt$.  
	Following the ref.\cite{Hart2008}, 
	{we start with the fixed metric of    AdS$_{d+1}$  blackhole,}  
	\begin{small}
		\begin{equation}
			ds^2= -f(r)dt^2+\frac{dr^2}{f(r)}+r^2d\vec{x}^2,  \;\;  f(r)=r^2\left(1-\frac{r_h^d}{r^d}\right).\label{eq:1}
		\end{equation}
	\end{small}
	In this letter, 	we will consider only $d=4$ for technical simplicity. 
	The AdS radius is set to be $1$ and $r_h$ is the radius of the horizon.  The   temperature is given by 
	$T=\frac{d}{4\pi} r_h $ as usual. 
 	 The field equations become  
		\begin{small}
			\bea
			&&  \frac{d^2 \Psi }{d z^2} -\frac{d-1+z^d}{z(1-z^d)}\frac{d \Psi }{d z}+\left( \frac{g^2 \Phi^2}{r_h^2(1-z^d)^2}-\frac{m^2}{z^2(1-z^d)}\right)\Psi =0, \nonumber \\
			&& \frac{d^2 \Phi}{d z^2}-  \frac{d-3}{z}\frac{d \Phi}{d z } -\frac{2g^2\Psi^2}{z^2(1-z^d)}\Phi =0.
			\label{eq:3}
			\eea
		\end{small}
		with the coordinate $z= r_h/r$. 
		One should notice that the regions $z>1$ and $0<z<1$ are inside and outside of the black hole respectively.  
		Here, $\Psi(z)$ is the scalar field  and the electrostatic scalar potential  $A_{t}=\Phi$.
	Near the boundary $z=0$, we have  
	\bea
	\Psi(z)&=& z^{\Delta_{-}}\Psi^{(-)}(z)+z^{\Delta_{+}}\Psi^{(+)}(z),
	\nonumber \\ \Phi(z) &=& \mu- ({\rho}/{r_h^{d-2}})z^{d-2} +\cdots
	\label{eq:4}
	\eea
	where $\Delta_{\pm}$  are related by 
	$\Delta_{+}+\Delta_{-}=d$ and   
	 $\mu$ and  $\rho$  are the chemical potential  and the charge density, respectively. 
	 Once $\Delta$ is determined,   $m^{2}$ follows using  $m^{2}=\Delta(\Delta-d)$. 
	We restrict ourself to  the near critical temperature where 
	probe solution can be trusted \cite{Horo2009}. 

	\vskip.2cm\section{ Near critical temperature }\label{hahaha}
  The critical temperature  is determined\cite{Siop2010} by the  the Sturm Liouville eigenvalue $\lambda$. 
In this section, we will  find the relation between   $\lambda$,   and the scaling dimension $\Delta$. This section is a brief review of our previous work \cite{Choun:2021pvs}. 
 
	At  the critical temperature $T_c$, $\Psi =0$, so  Eq.(\ref{eq:3}) tells us $\Phi^{''}=0$ near there. Then, we can set \cite{Siop2010} 
	\begin{equation}
		\Phi(z)= \tilde{\lambda}  r_c (1-x) \hspace{1cm}\mbox{where}\;\;\tilde{\lambda} =\frac{\rho}{r_c^3}
		\label{qq:9}
	\end{equation}
	where $x=z^2$ and $r_{c}$ is horizon radius at the critical temperature. For $T\rightarrow T_c$, the field equation $\Psi$ approaches to \cite{Choun:2021pvs}
	\begin{equation}
		-\frac{d^2 \Psi }{d x^2} +\frac{1+x^2}{x(1-x^2)}\frac{d \Psi }{d x}+  \frac{m^2}{4 x^2(1-x^2)} \Psi =\frac{\lambda ^2}{4x(1+x)^2}\Psi
		\label{qq:10}
	\end{equation}
	where $\lambda = g\tilde{\lambda}$. The critical temperature is given by  \cite{Siop2010,Choun:2021pvs}
	\be
	T_c 
	=\frac{d}{4\pi}r_c
	=\frac{1}{ \pi} \left(\frac{g \rho }{\lambda }\right)^{\frac{1}{3}},
	\label{si:1}
	\ee
	for $d=4$. 
	Factoring out the behavior near  	$x=0$ and 
	$x=-1$, we  have 
		\begin{equation}
		\Psi(x)= \frac{\left< \mathcal{O}_{\Delta}\right>}{\sqrt{2}r_h^{\Delta}}x^{\frac{\Delta}{2}}	(1+x)^{-\lambda /2}y(x) . 
		\label{qq:11}
	\end{equation} 



	Here,  $y$ is normalized by $y(0)=1$ and  we obtain
	\begin{small}
		\begin{eqnarray}
			&&\frac{d^2 y }{d x^2} + \left( \frac{\rho_0}{x}+\frac{\rho_1}{x-1}+\frac{\rho_2 }{x+1}\right) \frac{d y}{d x}  \nonumber\\
			&&+\left(\frac{w_0}{x}+\frac{w_1}{x-1}+\frac{w_2}{x+1}\right) y=0,
			\label{qq:12}
		\end{eqnarray}
	\end{small}
	\begin{equation}
\hbox{where }		\begin{cases} \rho_0=\Delta-1, \quad 
			\rho_1=1 , \quad 
			\rho_2=1-\lambda ,\cr 
			w_0=\frac{\lambda }{2}\left(-\Delta +\frac{\lambda }{2}+1\right), \cr
			w_1=\frac{1}{8}(\Delta ^2-2 \lambda ), \cr
			w_2=\frac{1}{8} \left(-\Delta ^2+4 \Delta  \lambda -2 \lambda ^2-2 \lambda \right).        
		\end{cases}
		\nonumber
	\end{equation} 
	Eq.(\ref{qq:12}) is the Heun's differential equation \cite{Ronv1995} that has four regular singular points at $x=0,1,-1,\infty$. 
	Substituting $y(x)= \sum_{n=0}^{\infty } d_n x^{n}$ at $|x|<1$ into (\ref{qq:12}), we obtain  a   three-term recurrence relation:
	\begin{equation}
		\alpha_n\; d_{n+1}+ \beta_n \;d_n + \gamma_n \;d_{n-1}=0,   \quad
		 \label{qq:13}
	\end{equation}
	 for   $n \geq 1$, with
	\begin{equation}
		\begin{cases} \alpha_n= (n+1)(n+ \Delta-1) \cr
			\beta_n= -\frac{\lambda }{2}\left(2n+\Delta-1-\frac{\lambda }{2}\right) \cr
			\gamma_n=-\left( n -1+\frac{\Delta}{2}-\frac{\lambda }{2}\right)^2 .        
		\end{cases}
		\label{qq:14}
	\end{equation}
	The first two  $d_{n}$'s are determined by   $\alpha_0 d_1+ \beta_0 d_0=0$ and  $d_{0}=1$,  the latter of which is due to the linearity of the equation. 
	
	%
	%
	
 Now we assume that the series converges at   $x=\pm 1$. For this, we introduce the concept of `minimum solution' : 	let  Eq.(\ref{qq:13})  $X(n)$, $Y(n)$ be the two linearly independent solutions for $d_{n}$.   $X(n)$ is called a minimal solution of  Eq.(\ref{qq:13}) if $\lim_{n\rightarrow\infty}X(n)/Y(n)=0$  and not all
	$X(n)=0$.  
	It has been known \cite{Ronv1995} that we have a convergent solution of $y(x)$ at $|x|=1$ if and only if the three term recurrence relation Eq.(\ref{qq:13}) has a minimal solution.   Eq.(\ref{qq:13}) has two linearly independent solutions $d_1(n)$, $d_2(n)$. One can show that  \cite{Jone1980} for large $n$, 
	\begin{equation}
		\begin{cases}  d_1(n)\sim n^{-1} , \cr
			d_2(n)\sim   (-1)^n  n^{-1-\lambda } , 
		\end{cases}
		\label{qq:32}
	\end{equation}
	which says  
	$	\lim_{n\rightarrow\infty} { d_2(n)}/{ d_1(n)}=0,$
	because $\lambda >0$. 
	Therefore  $d_2(n)$ is a minimal solution.  
	
	Now, we are in the position  to   calculate the $\lambda $. 
	According to Pincherle’s Theorem  \cite{Jone1980},
	$(d_n)_{n\in \mathbb{N}}$ is the minimal solution  if 	  the continued fraction 
	\bea 
		&& \beta_0 -\cfrac{\alpha_0 \gamma_1}{\beta_1 -\cfrac{\alpha_1 \gamma_2}{\beta_2-\cfrac{\alpha_2
					\gamma_3}{\beta_3 -\cdots}}} =0, \label{eq:19} \\
 %
 \hbox{ or, }
&& 	\beta_0 - \frac{\alpha_0 \gamma_1}{\beta_1 -}\frac{\alpha_1 \gamma_2}{\beta_2 -}\frac{\alpha_2 \gamma_3}{\beta_3 -} \cdots \frac{\alpha_{N-1} \gamma_{N}}{\beta_{N}}  = 0,
			\label{eq:24}
	\eea
		 for   sufficiently large $N$.
	One should remember that $\alpha_{n},\beta_{n},\gamma_{n}$'s are functions of $\lambda $
	so that eigenvalues  are the solution of the above equation.   Notice also that 
	Eq.(\ref{eq:24})   becomes a polynomial of degree $2N+2$ with respect to $\lambda $. 
	{	Therefore, the algorithm for finding $\lambda$ for a given $\Delta$ is as follows:    
		\begin{enumerate}
			\item  
			We substitute Eq.(\ref{qq:14}) into Eq.(\ref{eq:24}). 
			\item
			Choose a finite $N$ and workout 
			Eq.(\ref{eq:24}).  
			\item
			Find the zeroes of this equation.
			\item
			Increase $N$ until the root converges to a constant value within the desired precision \cite{Leav1990}.  
		\end{enumerate}  
		We find  their roots by calculating the continued fraction using Mathematica. 
	The result is given as the real line of the figure 
	1. 
		\begin{figure}[h]
		\centering
		\includegraphics[width=0.7\linewidth]{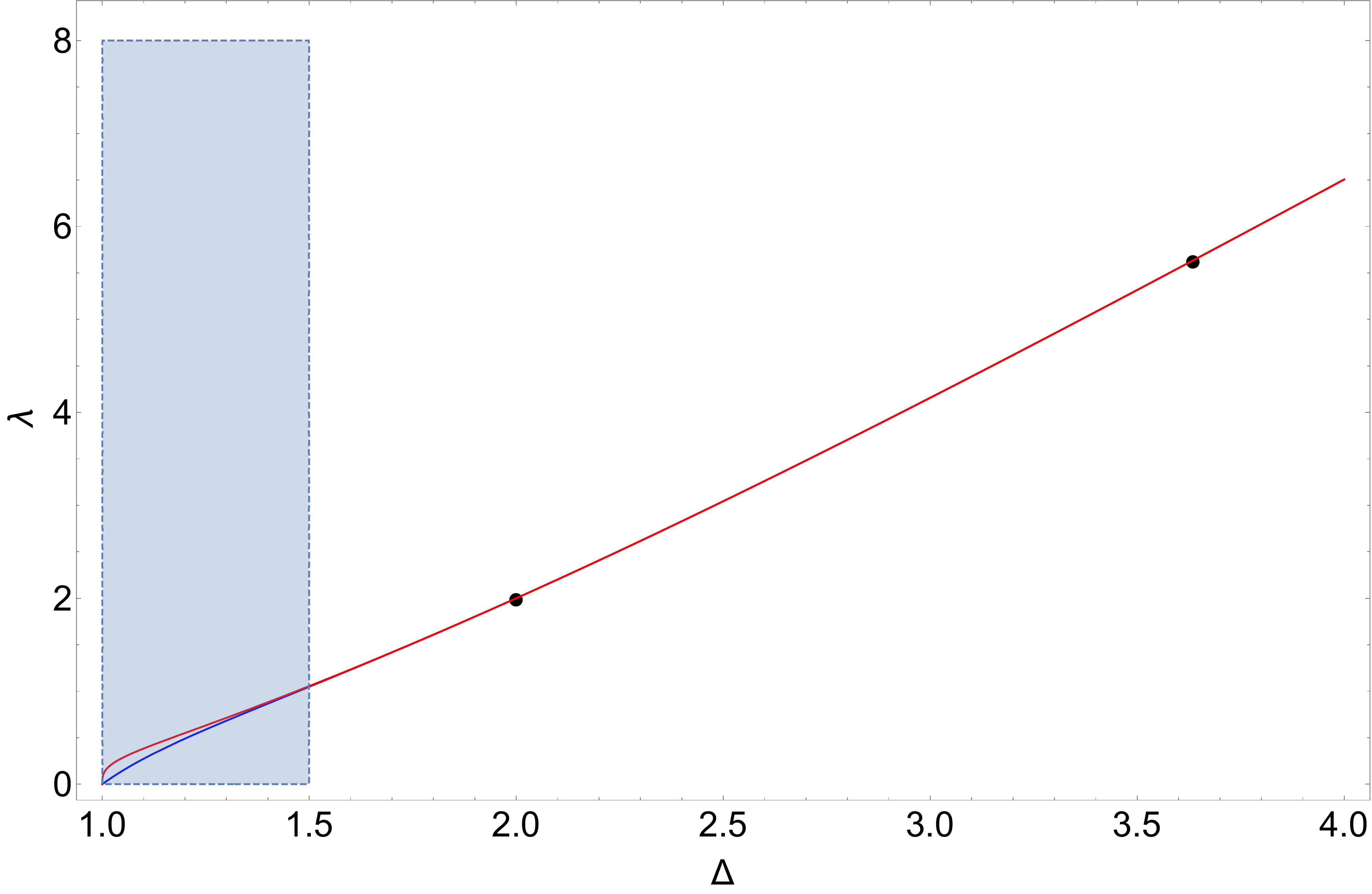}
		\caption{   $\lambda $  vs $\Delta$:  Blue and  red  colored curves are for $N=30, 31 $ respectively in  $\det\left(M_{N\times N}\right) =0$. 
		Two   dotted points are those given by the first two polynomial solution obtained by  eq. (\ref{Omega}) and eq. (\ref{app:1}). 
They are on the curve of values calculated by the Pincherle's theorem method, showing the consistency of the two calculations.  
		} 
		\label{bcs_tc}
	\end{figure} 
We are only interested in the smallest positive real root of $\lambda$.  We choose $N=30, 31$. 	
 Notice that there are two branches in  the shaded region, $1<\Delta<3/2$, which means that there is no well defined eigenvalues in this regime. 
	We  find  that a good fit for the numerical result can be given by 
		\be
	\lambda  \approx 1.18 \Delta^{4/3} - 0.97 \;\;\mbox{for}\;\; 3/2\leq \Delta\leq 4, \label{qq:36}
	\ee
	so that the eigenvalue is a continumous function of $\Delta$. 
The critical temperature  	can now be calculated by Eqs. (\ref{si:1}) and (\ref{qq:36}).
 On the other hand,  for $1<\Delta<3/2$ which is the shaded region in Figure 1,   $\lambda$ hence the critical temperature  is not well defined.  Therefore in this paper we  only consider the region $\Delta>3/2$. 
	
\vskip.2cm\section{ Scaling dimension from the black hole interior }\label{Tcc}  
Now we come back to our main goal, the determination of the discrete values of allowed scaling dimension. 

If we include  the interior of the black hole as well as outside as  the domain of  the Heun's equation, eq.(\ref{qq:12}) should be a polynomial, because eq.(\ref{qq:32}) shows that the infinite series is divergent at $x\geq 1$.  If the degree of the polynomial is $N$, then 
we need to impose 
\be
d_{N}\neq 0, \quad d_{N+1}=d_{N+2}=0,
\ee
  which is necessary and sufficient condition for the solution to be a degree $N$ polynomial. The equation (\ref{qq:13})
request that $\gamma_{N+1}=0$ should hold as well. 
Then,   there are essentially two conditions for which we need  to impose 
\be
\gamma_{N+1}=d_{N+1}=0 \quad  \mbox{ for degree  } N \in \mathbb{N}_{0},  \label{aa:1}
\ee 
because in this case $d_{N+2}=0$ iff $\gamma_{N+1}=0$ under the assumption of $d_{N+1}=0$.  Notice that since all $d_{n}$ are functions of the parameters  in  the differential equation, there should be at least two parameters which can be fine tuned to satisfy above two conditions. This means that in our case there are two parameters which should be quantized. We call them 'eigenvalues'. We remind the readers  that for the hypergeometric case which has only three singularities at $0,1,\infty$,  the recurrence  equations involve only two terms ($d_{n}$ , $d_{n+1}$) after factoring out the solution's behaviors near the  singularities at the zero and infinity, and we  only  need to impose $d_{N+1}=0$ which   gives us   quantization of one parameter, the energy in Schroedinger equation for example.   
For the system with more than three singularities, we   meet three or more term recurrence relation, which is our case. 

Now coming back to our case, 
if the equations  contains exactly two parameters,      they are generically quantized, because   the solutions corresponds to the  intersection points of the two curves defined by 
eqs.(\ref{aa:1}).
In our case, we have $\lambda$ and $\Delta$ and these parameters are quantized.  	More explicitly, from eq.(\ref{qq:14}), 
$B_{N+1}=0$ gives 
\be
\lambda =2N+\Delta. \label{Omega}
\ee 
One  interesting consequence of this result  is that our solution of scalar field given in \eqref{qq:11}  always has asymptotic behavior which saturate to the finite  constant. That is, although $y$ is a polynomial, the 
scalar function itself has  well defined asymptotic value at $z\to\infty$.. It happened to be finite although we never requested its finiteness. 
 \begin{equation}
		\Psi(x\to \infty)  
		\simeq  \frac{\left< \mathcal{O}_{\Delta}\right>}{\sqrt{2}r_h^{\Delta}}.
		\label{qq:111}
	\end{equation}
We plot  solutions $\Psi(x)$ for $N=0,1,2$ and 30 in Fig.~\ref{polynomial}. 
    In fact, the issue of the  solution  of holographic superconductor inside black hole was studied in recent paper by Hartnoll et.al.  \cite{Hartnoll:2020fhc}
 from the different perspective. 
 Our result corresponds to the solution to the 
 linearized level. Nevertheless, the oscillation and its death are the same features for large polynomial order.  

 		\begin{figure}[!htb]
		\centering
			\subfigure[]
		{ \includegraphics[width=0.6\linewidth]{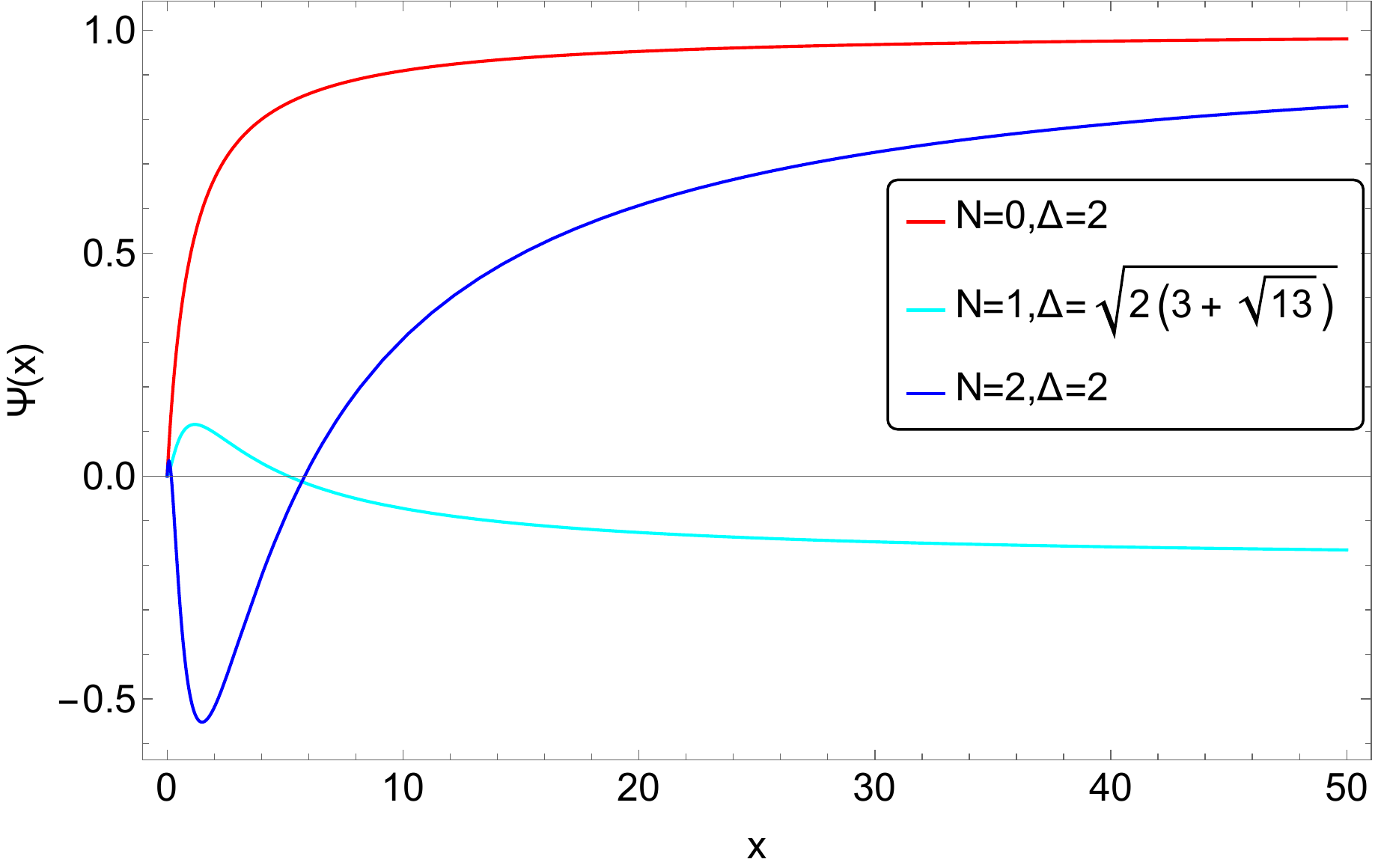}}
	\subfigure[]
	{ \includegraphics[width=0.6\linewidth]{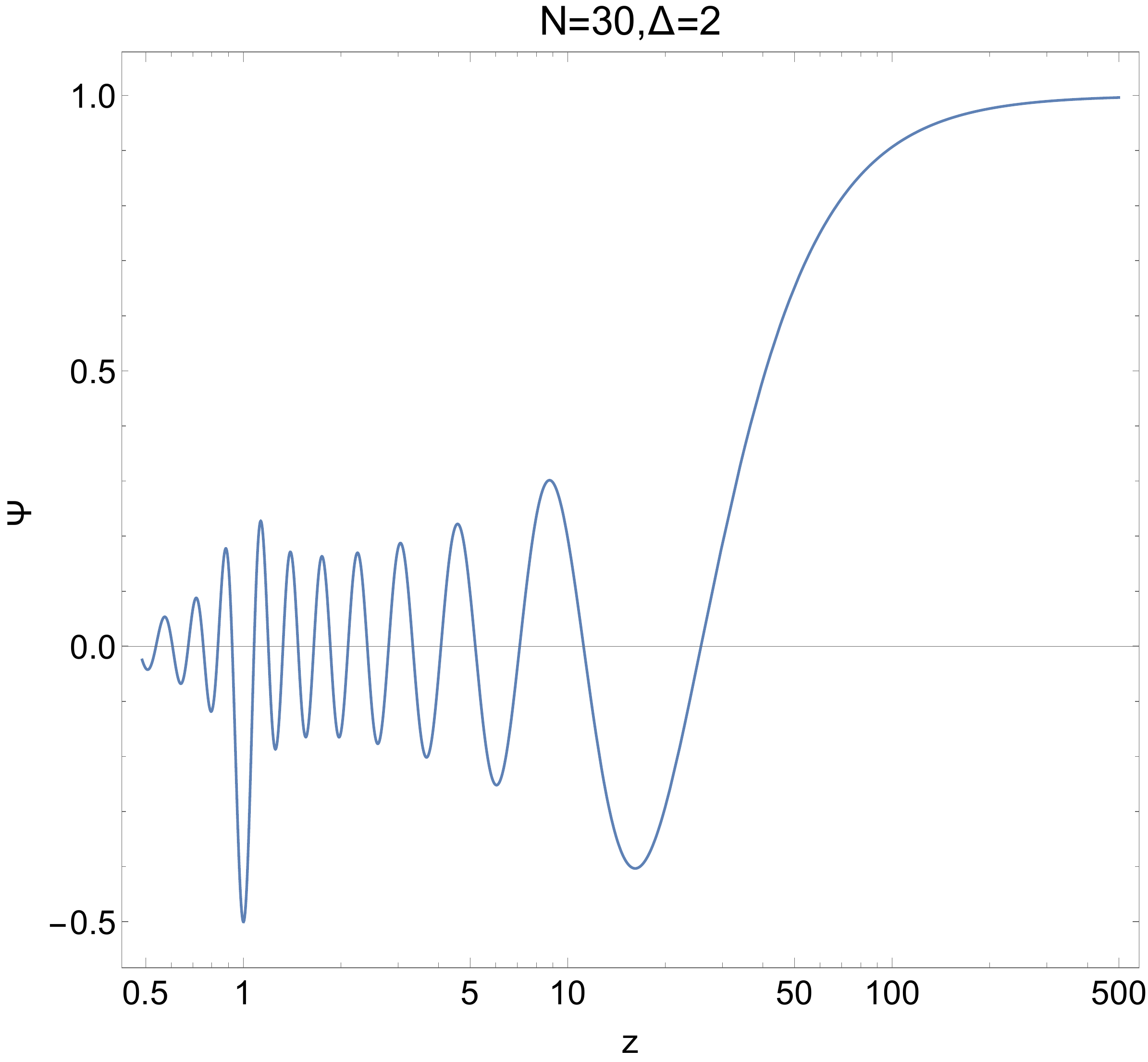}}  
		\caption{ $\Psi(x)$ for $N=0,1,2$. Here, we set $\frac{\left< \mathcal{O}_{\Delta}\right>}{\sqrt{2}r_h^{\Delta}}=1$ for convenience. And $\Psi(x\to \infty) $ is finite constant.	(b)  N=30.	} 
		\label{polynomial}
	\end{figure} 

Now, $d_{N+1}=0$  gives  a $N+1$-th order polynomial  in $\Delta$, which we call $ {\cal P}_{N+1} $,  so that
$	{\cal P}_{N+1}(\Delta)=0$. 	 
    Low-order expressions of  these polynomials are given by
{\small
	\bea
	\label{app:1}
	\begin{split} {\cal P}_{1}(\Delta)&=  \Delta(\Delta-2) ,\\
		{\cal P}_{2}(\Delta)&=  \Delta^4 -12 \Delta^2-16  ,\\
		{\cal P}_{3}(\Delta)&=  (\Delta-2)(\Delta+4)(\Delta^5+2\Delta^4-36\Delta^3-8\Delta^2+256)  .
	\end{split}
	\eea
}
These  	tell us that 
\begin{description} 
	\item[$N=0$] ~~
	$ \Delta=2,  \hbox{ and } \lambda =2,   $
	\item[$N=1$ ]  ~~
	$\Delta=\sqrt{6+2\sqrt{13}},   \hbox{ and } \lambda =2+\Delta.  $  
	\item[$N=2$ ]  ~~
	$\Delta= 2,\sqrt{8 \sqrt{6}+17}-1,   \hbox{ and } \lambda =4+\Delta.  $  
\end{description}
	\begin{figure}[h]
	\centering
	\includegraphics[width=0.7\linewidth]{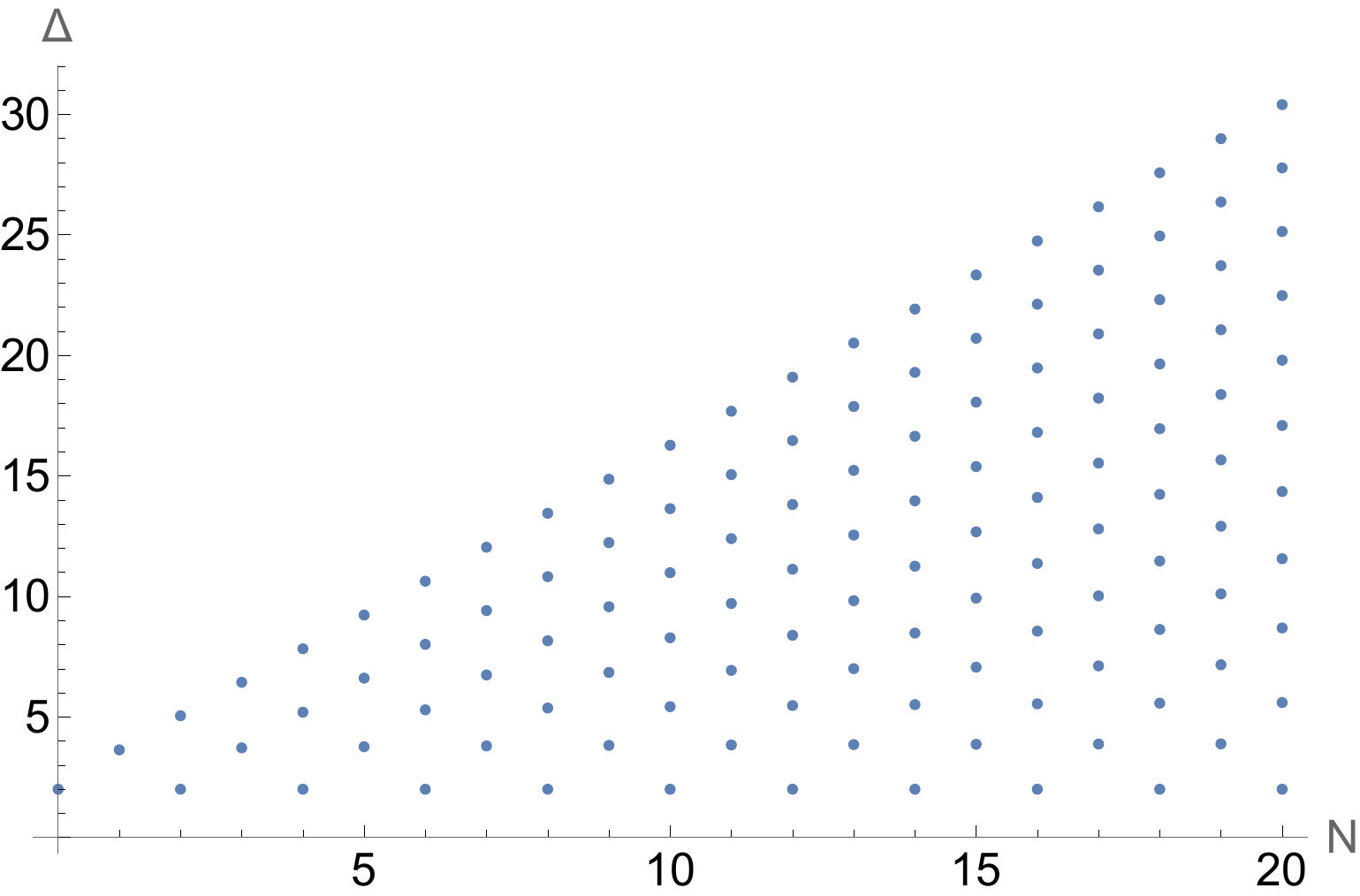}
	\caption{   Allowed $\Delta $  vs $N$.   	} 
	\label{deltalists}
\end{figure} 
 Fig.~\ref{bcs_tc}	shows us that above allowed values  $( \Delta, \lambda )= (2,2)$ and $ (3.635, 5.635)$ as $N=0, 1$ are placed on  the line  of the  $\Delta, \lambda$, which would be  obtained by Pincherle's method when   we request that the solution is well defined only outside the black hole.   We remark that we did not set $m^{2}_{\Phi}=-2$  to get  $\Delta=2$. 
Our method  can be regarded as a calculational tool for $ \Delta$. 
Also, notice that 
on the allowed   points are on the curve  
obtained in the previous section. See the   black dots in Fig. 1. Fig.~\ref{deltalists} shows us all $\Delta$'s up to $N=20$.  Due to the relation \eqref{Omega},  lower $\Delta$ and lower $N$ solotions are more stable under the perturbation since they give lower eigenvalue $\lambda$. 
 
 \section{  Regularity  conditions \label{FDE}  }\
In the presence of the black hole, we often imposes contraints  by requesting that the differential equation is well defined at the horizon. Then it is an urgent question whether such regularity constraints imposes further quantization condition. We will show below that this is not the case.

	We consider the differential equation such as	 
	\begin{equation}		
		E''(z)+ \sum _{j=0}^k \frac{\rho _j}{z-b_j} E'(z) +\sum _{j=0}^k \frac{w_j}{z-b_j}	E(z) =0 , \label{ff:2}
	\end{equation} 	with $ \sum _{j=0}^k w_j =0$.  We set   $b_0=0$ for the ease of the analysis. The case   $k=2$ is the Heun's equation.
	The regularity conditions at three singularities at finite positions are 
	\begin{equation}
		\begin{cases} b_1 b_2 \left(\rho _0 E'(0)+E(0) w_0\right)=0, \cr
			b_1 \left(b_1-b_2\right) \left(\rho _1 E'\left(b_1\right)+w_1 E\left(b_1\right)\right)=0,\cr 
			b_2 \left(b_2-b_1\right) \left(\rho _2 E'\left(b_2\right)+w_2 E\left(b_2\right)\right)=0.        
		\end{cases}
		\label{ff:3}
	\end{equation}  
	The solution of Eq.(\ref{ff:2}) is expressible by a  Frobenius series. According to Fuchs' theorem, its radius of convergence is at least as large as the minimum  of the radii of convergence of $\sum _{j=0}^k \frac{\rho _j}{z-b_j}$ and $\sum _{j=0}^k \frac{w _j}{z-b_j}$. 
	If we require that the domain of a solution of Eq.(\ref{ff:2}) is entire complex plane or real line, 
	the solution  should be a polynomial. 
	Suppose it is  of degree  $N$.	
	After factoring out the behavior near $z=\infty$ and dividing  the Eq.(\ref{ff:2}) by $E''(z)$, 
	the following   is the leading terms near $z=\infty$:       
	\begin{small}
		\begin{equation}
			\Big(\sum _{j=1}^2 b_j w_j+N \big(\sum _{j=0}^2 \rho _j+N-1\big)\Big)+z \sum _{j=0}^2 w_j =0. 	\label{ff:5}
		\end{equation}
	\end{small}
	The  vanishing of the second term is the regularity condition which was already required in the definition of the Fuchsian  equation. The first term 
	  requests :		\begin{equation}
			\sum _{j=1}^2 b_j w_j+N \Big(\sum _{j=0}^2 \rho _j+N-1\Big)=0, 
		\label{ff:6}
	\end{equation} 
which  is the condition for the solution to be a polynomial of degree $N$. 
	From Eq.(\ref{ff:3}) and Eq.(\ref{ff:6}), 
	we have 4 conditions to be satisfied.  
  One may worry that  the problem could be over determined and  in general we might not have a solution. So 
	our question is how many of these regularity conditions are automatically satisfied due to the equation of motion.  We will prove that 
	{   all regularity conditions are satisfied automatically by the solution of equation of motion. }
	Therefore the regularity condition will not request any further constraint. 
	For this we repeat the calculation in slightly more general setting. 
		
	Let $E(z) =\sum_{n=0}^{\infty} d_n z^n$ and substitute it into Eq.(\ref{ff:2}).
	\begin{small}
		\begin{equation}
			\sum _{n=0}  \alpha _n d_{n+1}   z^n+\sum _{n=0}  \beta _n d_n   z^n+\sum _{n=1}  \gamma _n d_{n-1}   z^n=0 	\label{fff:1}
		\end{equation}
	\end{small}
	with
	\begin{small}
		\begin{equation}
			\begin{cases}\alpha _n=b_1 b_2 (n+1) \left(n+\rho _0\right) ,\cr
				\beta _n = w_0 b_{,1}^2-n \left((n-1) b_{,1}^2+b_{,1}^2 \rho _{,0}^2-(b\rho)_{,1}^2\right) , \cr
				\gamma _n= \left(b_1 w_1-b_2 w_{,0}^1\right)+(n-1) \left(\rho _{,0}^2+n-2\right) ,  
			\end{cases}
			\label{fff:2}
		\end{equation}
	\end{small}
	where $b_{,1}^2=\sum _{j=1}^2 b_j$, $\rho _{,0}^2=\sum _{j=0}^2 \rho _j$, $(b\rho)_{,1}^2=\sum _{j=1}^2 b_j \rho _j$ and $w_{,0}^1=\sum _{j=0}^1 w_j$.
	Then eq.(\ref{fff:1}) becomes 
	\begin{small}
		\begin{eqnarray}
			&&\alpha _0 d_{1} +\beta _0 d_0  +	\sum _{n=1}^{\infty}  \left( \alpha _n d_{n+1} + \beta _n d_n +\gamma _n d_{n-1} \right)  z^n  =0.
 			\label{fff:3}
		\end{eqnarray}
	\end{small}	 
	As   before,  for the series to teminate at $d_{N}z^{N}$, we need 
 	\bea
	\alpha _0 d_{1} +\beta _0 d_0&&=0, \cr
	\alpha _n d_{n+1} + \beta _n d_n +\gamma _n d_{n-1}&&=0,  
	\hbox{ for~ all } 1\leq n   \cr
	d_{N+1}=0,\quad \gamma_{N+1}&&=0. \label{rec}
	\eea
With these,  the LHS of the first regularity  condition of  eq.(\ref{ff:3}) is 
\be 
b_1 b_2 \left(\mathit{d}_1 \rho _0+\mathit{d}_0 w_0\right), \ee
 which vanishes by the first relation of eq.(\ref{rec}). 
The LHS of the second regularity  condition of  eq.(\ref{ff:3}) becomes
\begin{small}
	\begin{equation}
		 \sum _{n=1}^{\infty}  \left(\tilde{\alpha }_n\mathit{d}_{n+1}  +\tilde{\beta }_n\mathit{d}_n  +\tilde{\gamma }_n\mathit{d}_{n-1}  \right) b_1^n
		  +b_1 b_2 \left(\mathit{d}_1 \rho _0+\mathit{d}_0 w_0\right),    \label{regul}
	\end{equation}
\end{small}
with
\begin{small}
\begin{equation}
	\begin{cases} \tilde{\alpha }_n= \alpha _n-b_1 b_2 n (n+1), \cr
	\tilde{\beta }_n =\beta _n+\left(b_1+b_2\right) (n-1) n,
		\cr 
		\tilde{\gamma }_n= \gamma _n-(n-2) (n-1).        
	\end{cases}
	\nonumber
\end{equation} 
\end{small}	
All terms in  eq. (\ref{regul}) vanish due to the eq. (\ref{rec}). We can easily check 
$\tilde{\alpha }_n$, 	$\tilde{\beta }_n$ and $\tilde{\gamma }_n$ have the following relation  
\begin{equation}
	 \tilde{\alpha }_n+ \tilde{\beta }_{n+1}b_1 +  \tilde{\gamma }_{n+2}b_1^2  = \alpha _n+  \beta _{n+1}b_1 +  \gamma _{n+2} b_1^2.
	\nonumber
\end{equation} 	
	Now, notice that 
\bea
&&\sum _{n=1}  \left(\tilde{\alpha }_n\mathit{d}_{n+1}  +\tilde{\beta }_n\mathit{d}_n  +\tilde{\gamma }_n\mathit{d}_{n-1}  \right) b_1^n\\ 
&&=\sum _{n=1} b_1^n \mathit{d}_{n+1} \left(\tilde{\alpha }_n+  \tilde{\beta }_{n+1}b_1 +  \tilde{\gamma }_{n+2}b_1^2 \right),\nonumber\\
&&+b_1 \mathit{d}_1 \tilde{\beta }_1+b_1^2 \mathit{d}_1 \tilde{\gamma }_2+b_1 \mathit{d}_0 \tilde{\gamma }_1\\
&&=\sum _{n=1} b_1^n \mathit{d}_{n+1} \left( \alpha _n+   \beta _{n+1}b_1 + \gamma _{n+2} b_1^2\right), \nonumber\\
&&+\beta _1 b_1 \mathit{d}_1+b_1^2 \gamma _2 \mathit{d}_1+b_1 \gamma _1 \mathit{d}_0,\\
 	&&=\sum _{n=1}   \left(\alpha _n \mathit{d}_{n+1}+\beta _n \mathit{d}_n+\gamma _n \mathit{d}_{n-1}\right)b_1^n =0 \\
	\nonumber	\eea 
	 which vanishes by the recurrence relation  in  eq(\ref{rec}). 	
	 The regularity conditions at $z= b_2$ in eq.(\ref{ff:3})  is  satisfied  by 
	 $b_{1}\to b_{2}$. 
	Finally the  second equation of eq.(\ref{ff:6}) is equivalent to $\gamma_{N+1}=0$ as one can see from the expression in eq.(\ref{fff:2}). 
	Therefore, {\it all the 4 regularity conditions are automatically satisfied by the polynomial  solutions of the equation. }

	\vskip.2cm
	\section{Discussion}
	Our work is   for AdS$_{5}$ dual to a 3+1 dimensional system. 
	For  AdS$_{4}$ blackhole, 	we  have a technical difficulty in applying our method: while   AdS$_{5}$ metric is even under the $z\to -z$ reflection,  we do not have such symmetry in AdS$_{4}$. Therefore  we can not reduce the singularity of the differential equation.   As a consequence, the Heun's equation leads us to a four term recurrence relation. In this case, for a solution to be valid inside the black hole,  we need at least 3 parameters
	while we have only two. 
	
We also would like to mention the key difference from the previous literatures. While the previous solutions request just the regularity of the solution near the horizon, we claim that the horizon regularity condition implies  the regularity at the  center of black hole \cite{YoonSeokChoun}.

	\vspace{ 0.2cm}
	\acknowledgments
	This  work is supported by Mid-career Researcher Program through the National Research Foundation of Korea grant No. NRF-2021R1A2B5B02002603,   NRF-2020-R1A2C2-007930,  NRF-2022H1D3A3A01077468. 
	We also  thank the APCTP for the hospitality during the focus program, “Quantum Matter and   Entanglement with Holography”, where part of this work was discussed.
	
	

	\bibliographystyle{JHEP}
	\bibliography{Refs_SC}

\end{document}